\documentclass[sigconf,nonacm]{acmart}

\AtBeginDocument{%
  }

\usepackage{graphicx}
\usepackage{booktabs}
\usepackage{multirow}
\usepackage{xcolor}
\usepackage{subcaption}
\usepackage{enumitem}
\usepackage{float}
\usepackage{dsfont}

\setcopyright{none}
\copyrightyear{2026}
\renewcommand\footnotetextcopyrightpermission[1]{}

\acmYear{2026}

\begin{document}
\title{Proactive Dimension-Adaptive Denoising for Graph Collaborative Filtering via Representation Coherence}


\author{Wenjun Li}
\authornote{Equal contribution.}
\affiliation{%
  \institution{Netease Media Group}
  \city{Beijing}
  \country{China}
}
\email{liwenjun01@corp.netease.com}

\author{Dong Zhao}
\authornotemark[1]
\authornote{Corresponding author.}
\affiliation{%
  \institution{Netease Media Group}
  \city{Beijing}
  \country{China}
}
\email{zhaodong03@corp.netease.com}

\author{Jiaying Shi}
\affiliation{%
  \institution{Netease Media Group}
  \city{Beijing}
  \country{China}
}
\email{shijiaying@corp.netease.com}

\author{Juan Zhang}
\affiliation{%
  \institution{Netease Media Group}
  \city{Beijing}
  \country{China}
}
\email{zhangjuan1@corp.netease.com}

\author{Zhaoming Pan}
\authornotemark[2]
\affiliation{%
  \institution{Netease Media Group}
  \city{Beijing}
  \country{China}
}
\email{panzhaoming@corp.netease.com}

\author{Zhimin Zhang}
\affiliation{%
  \institution{Netease Media Group}
  \city{Beijing}
  \country{China}
}
\email{zhangzhimin@corp.netease.com}

\renewcommand{\shortauthors}{Li et al.}

\begin{abstract}
Implicit feedback in recommender systems contains noisy interactions (e.g., accidental clicks, impulse purchases) that propagate through user--item graphs and contaminate learned representations. Existing explicit denoising methods uniformly rely on training loss to estimate interaction reliability. However, loss is a fundamentally passive signal: it detects noise only after it has caused prediction errors, and it is heavily confounded by training stage, negative sampling, and hard positives.

We propose \textbf{Noise-aware Dimension Learning (NDL)}, a proactive dimension-adaptive denoising framework. NDL introduces \textbf{Representation Coherence Detection (RCD)}, which evaluates the coherence between an interaction representation and the user's historical preference centroid in embedding space, providing an active noise signal independent of loss. RCD and the loss signal are fused via activity-aware weighting (active users rely more on coherence while cold-start users automatically fall back to loss), and the fused estimate drives an interaction-level dimension-adaptive mask that limits the effective representation capacity of unreliable interactions. A lightweight contrastive regularizer further improves embedding uniformity, stabilizing the geometry on which RCD discriminates. On three public benchmarks NDL consistently outperforms state-of-the-art denoising methods, lifting Recall@5 by \textbf{+17.95\%} on Amazon-book, \textbf{+12.33\%} on Yelp, and \textbf{+5.63\%} on MovieLens-1M over the strongest baseline DAR; the core variant without contrastive learning already surpasses DAR on sparser datasets, confirming representation coherence as the primary driver of these gains. NDL is confined to the training loop and is serving-compatible with LightGCN-based production pipelines, adding no online latency or parameters at inference time. Code: \url{https://github.com/GordonDongZHAO/NDL}
\end{abstract}

\ccsdesc[500]{Information systems~Recommender systems}

\keywords{Dimension-Adaptive Learning; Implicit Feedback Denoising; Graph-Based Collaborative Filtering; Denoising Recommendation}

\maketitle

\section{Introduction}
\label{sec:intro}

Recommender systems are the backbone of modern online platforms, spanning e-commerce, streaming services, and social networks. Modern collaborative filtering~\cite{hu2008implicit,rendle2009bpr,he2017ncf} learns user and item representations from large-scale implicit feedback. Graph-based collaborative filtering methods, particularly LightGCN~\cite{he2020lightgcn}, have become the dominant paradigm by propagating embeddings on user--item bipartite graphs to capture high-order collaborative signals. However, real-world implicit feedback is inherently noisy: users may click out of curiosity, make impulse purchases during flash sales, or accidentally interact with irrelevant items. When these noisy interactions are involved in graph propagation, user preference representations are contaminated, leading to degraded recommendation accuracy and personalization.

\subsection{Limitations of Existing Denoising Methods}
\label{sec:limitations}

A growing body of work addresses noisy interactions from various perspectives. Crucially, we observe that virtually all existing explicit denoising methods share the \emph{same} noise estimation information source: the model's training loss.

Truncated loss (T-CE)~\cite{wang2021denoising} drops interactions whose loss exceeds a fixed threshold from gradient updates; Reweighted loss (R-CE)~\cite{wang2021denoising} assigns smaller gradient weights to interactions with higher loss; and Dimension-adaptive (DAR)~\cite{lai2025dar} maps per-interaction loss values to effective dimension counts.

Although these methods differ in \emph{how} they suppress noise, they are unified in \emph{where} the noise signal originates: cross-entropy (CE) loss. This is fundamentally a \emph{passive} mechanism---noise can only be detected \emph{after} the model has already been misled by noisy samples, manifested as large prediction errors. Moreover, CE loss is itself influenced by negative sampling stochasticity, user activity distributions, and training stage, making it unreliable as the sole noise indicator.

\subsection{Active Signals in Embedding Space}
\label{sec:insight}

Beyond the passivity issues above, training-stage ambiguity, popularity confounding, and late-training memorization~\cite{zhang2017rethinking,arpit2017memorization} further erode loss-based signals. We identify a complementary cue overlooked by prior work: in the embedding space of graph collaborative filtering, the interaction representation of a noisy interaction deviates from the user's aggregated history, showing markedly lower geometric coherence than clean ones.

\paragraph{Preference Coherence Hypothesis.} Let $\mathbf{r}_{ui} = \mathbf{e}_u \odot \mathbf{e}_i$ denote the interaction representation and $\bar{\mathbf{r}}_u = \frac{1}{|\mathcal{N}_u|} \sum_{j \in \mathcal{N}_u} \mathbf{e}_u \odot \mathbf{e}_j$ the user's historical mean. We posit:
\begin{align}
    \text{Clean:} \;\; \text{coh}_{ui} &= \cos(\mathbf{r}_{ui}, \bar{\mathbf{r}}_u) \approx 1, \\
    \text{Noisy:} \;\; \text{coh}_{ui} &= \cos(\mathbf{r}_{ui}, \bar{\mathbf{r}}_u) \ll 1.
\end{align}
Intuitively, a user's history forms a stable ``preference centroid'' $\bar{\mathbf{r}}_u$; an accidental click on an off-taste item deviates from it. The hypothesis assumes short-term preference stability---genuine exploration (e.g., new users) may also yield low coherence, a risk NDL mitigates via activity-aware fusion (\S\ref{sec:fusion}).

\begin{figure}[H]
    \centering
    \includegraphics[width=\columnwidth]{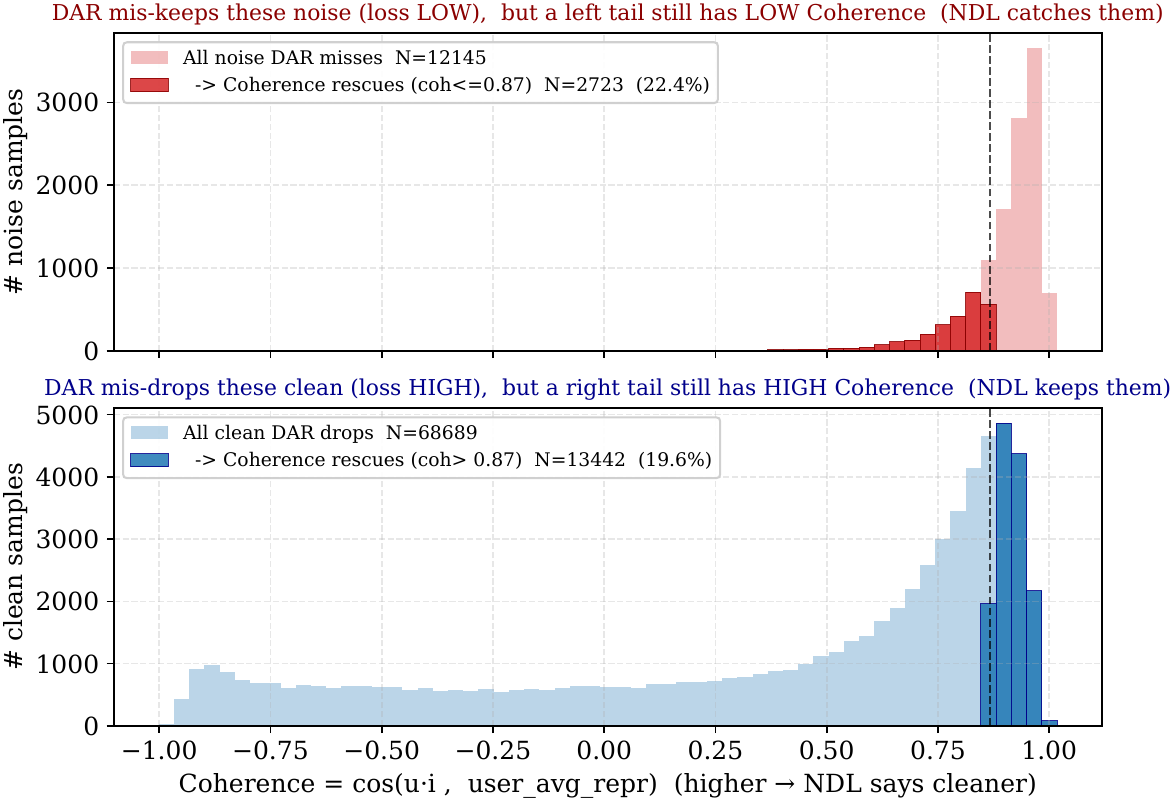}
    \caption{The ``rescue effect'' of coherence on subsets where loss-based discrimination fails. \textbf{Top}: false negatives (true noise with low BCE), $22.4\%$ have $\text{coh}\leq 0.87$. \textbf{Bottom}: false positives (clean with high loss), $19.6\%$ have $\text{coh}>0.87$.}
    \Description{Two stacked histogram panels comparing coherence distributions on the two error subsets where loss-based DAR fails. Top: coherence of false negatives (12,145 noisy samples retained by DAR), with a left tail where ~22.4\% fall below 0.87. Bottom: coherence of false positives (68,689 clean samples suppressed by DAR), skewed high, with ~19.6\% above 0.87. Together they show coherence recovers noise the loss misses and preserves clean samples the loss wrongly suppresses.}
    \label{fig:rescue}
\end{figure}

Figure~\ref{fig:rescue} quantifies this complementarity on MovieLens using the best DAR checkpoint, with BCE thresholded by DAR's adaptive cutoff and coherence by the sample median ($\approx 0.87$). On the two failure subsets of loss-based judgment, a non-trivial fraction of \emph{retained noise} still shows low coherence (deviating from the centroid despite well-fitted loss), while a comparable fraction of \emph{suppressed clean} samples retains high coherence (structurally aligned with long-term preferences). Thus loss and coherence are orthogonally complementary—loss captures immediate fitting, coherence reflects long-term preference conformity, free from training-stage, negative-sampling, or hard-positive ambiguities. This empirically grounds NDL's fusion of loss with coherence and explains why RCD simultaneously cuts false negatives and false positives, consistently beating DAR.

\subsection{Proposed Framework and Contributions}
\label{sec:contributions}

Rather than chasing a single ``perfect'' noise detector, NDL accepts that every signal has blind spots and cancels them out by fusing an active coherence signal with the passive loss signal, with a contrastive regularizer shaping the embedding geometry on which coherence is measured. Our contributions are:
\begin{itemize}[leftmargin=*]
    \item \textbf{Active coherence signal (RCD)}: an embedding-geometry-based noise estimator independent of prediction error, addressing the passivity, training-stage, and hard-positive ambiguity of loss-based denoising.
    \item \textbf{Activity-aware dimension-level denoising}: an activity-adaptive RCD/Loss fusion converted into interaction-level dimension-adaptive masks, so that reliable interactions use full capacity while unreliable ones are confined to limited dimensions.
    \item \textbf{Systematic validation}: on three benchmarks of widely different sparsity, NDL consistently improves over the SOTA baseline DAR; the core variant without contrastive regularization already surpasses DAR on the sparser Yelp and Amazon-book, isolating RCD itself as the dominant active signal.
\end{itemize}

\section{Related Work}
\label{sec:related}

\begin{figure*}[!t]
    \centering
    \includegraphics[width=\textwidth]{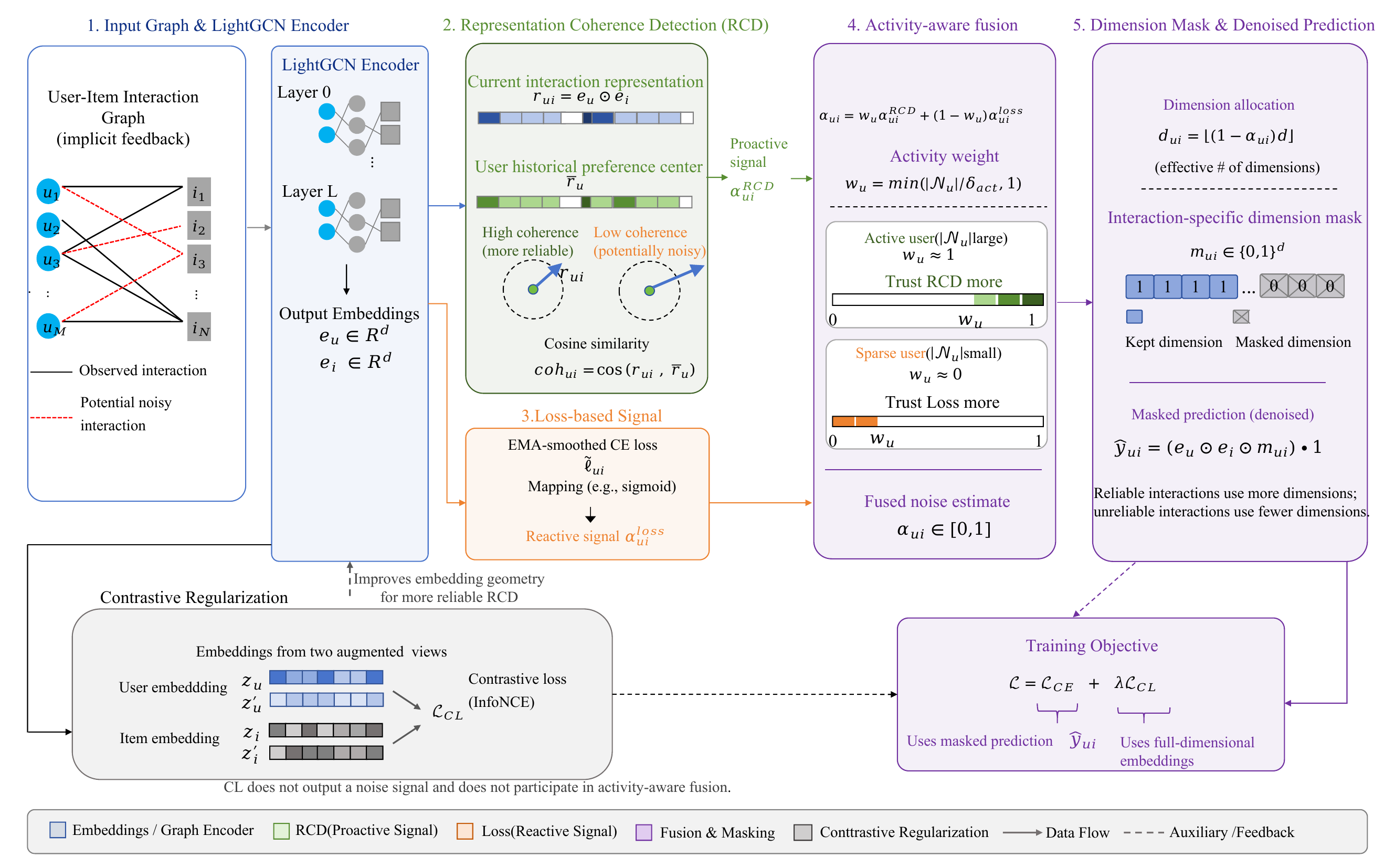}
    \Description{Schematic overview of the NDL training pipeline. The user--item graph is fed into a LightGCN backbone to produce user and item embeddings. From the embeddings, two parallel branches estimate per-interaction noise: an RCD branch that computes the cosine coherence between the current interaction representation and the user's historical preference centroid, and a Loss branch that uses an EMA-smoothed cross-entropy loss. The two noise signals are combined through an activity-aware fusion module weighted by user interaction count, and the fused noise estimate is converted into an interaction-level dimension-adaptive mask that controls how many embedding dimensions each interaction contributes to the prediction. A separate contrastive regularization branch perturbs the embeddings to produce two augmented views and applies an InfoNCE loss, serving only to improve embedding space geometry rather than to estimate noise.}
    \caption{Overview of the NDL framework. Two parallel branches estimate per-interaction noise (RCD on embedding geometry, Loss on EMA-smoothed CE); activity-aware fusion blends them; the fused score drives a dimension-adaptive mask that touches the prediction. CL is an off-path regularizer that only shapes embedding geometry.}
    \label{fig:framework}
\end{figure*}

\subsection{Graph Collaborative Filtering}

NGCF~\cite{wang2019neural} first introduced graph neural networks into collaborative filtering by performing message passing on user--item graphs. LightGCN~\cite{he2020lightgcn} simplified NGCF by removing feature transformations and nonlinear activations, retaining only neighborhood aggregation. LightGCN has become the de facto standard backbone for subsequent work and serves as the base model in our framework.

\subsection{Explicit Denoising for Recommendation}

Explicit denoising methods identify and suppress noisy interactions during training, and can be grouped by their noise signal source. \emph{Loss-based} methods (T-CE, R-CE~\cite{wang2021denoising}, BOD~\cite{wang2023bod}, Double Correction~\cite{he2024doublecorrection}, DAR~\cite{lai2025dar}) truncate, reweight, or dimension-adapt interactions based on per-sample loss; DAR in particular grounds the dimension mapping in the Information Bottleneck principle and achieves the state-of-the-art performance. \emph{Cross-model agreement} methods (DeCA~\cite{wang2022deca}, SGDL~\cite{gao2022self}) leverage two models or memorized clean signal to denoise. \emph{Graph-structural} methods (RGCF~\cite{tian2022rgcf}, SLED~\cite{zhang2023sled}, DDRM~\cite{zhao2024ddrm}) prune unreliable edges, transfer self-supervised structural priors, or apply diffusion-based embedding robustification. Other directions include reinforcement-learning instance selection (AutoDenoise~\cite{lin2023autodenoise}), preference-guided social pruning~\cite{quan2023robustpref}, willingness--action decomposition (UDT~\cite{chua2024unified}), and most recently LLM-based reasoning combined with information-bottleneck alignment (LLaRD~\cite{wang2025LLaRD}). This line of work is also conceptually related to noisy-label robust learning~\cite{zhang2018gce,han2018coteaching,li2020dividemix}.

A key observation is that the noise estimation of all above methods directly or indirectly depends on training objective feedback. The RCD proposed in this paper provides a noise signal from embedding geometric coherence, independent of prediction error, and orthogonally complementary to the above methods. Therefore, we adopt DAR's dimension-adaptive mechanism, originally derived from an information-bottleneck view~\cite{lai2025dar,tishby2000ib,tishby2015dl}, but drive it with our active coherence signal rather than DAR's loss-based score.

\subsection{Self-Supervised, Alignment, and Implicit Robustness}

The alignment--uniformity framework~\cite{wang2020understanding,wang2022directau} establishes that high-quality representations should keep similar entities close while spreading the embedding space uniformly; our RCD fits this view by detecting noise via misalignment to the user's historical pattern, and dimension adaptation prevents noise from contaminating high-value dimensions. Self-supervised contrastive methods (SGL~\cite{wu2021sgl}, NCL~\cite{lin2022ncl}, SimGCL~\cite{yu2022simgcl}, XSimGCL~\cite{yu2023xsimgcl}, HCCF~\cite{xia2022hccf}, LightGCL~\cite{cai2023lightgcl}; survey~\cite{yu2023sslsurvey}) and DirectAU~\cite{wang2022directau} improve recommendation through augmented views, and InfoNCE~\cite{oord2018cpc} implicitly promotes uniformity~\cite{yu2022simgcl,wang2020understanding} for some auxiliary noise robustness. However, this robustness is \emph{implicit}: CL methods cannot identify \emph{which} interactions are noisy nor apply differentiated per-interaction treatment. NDL treats the two paradigms as complementary, using a lightweight CL module solely to improve embedding uniformity for RCD's coherence-based noise detection.

\section{Method: Noise-Aware Dimension Learning}
\label{sec:method}

Figure~\ref{fig:framework} sketches the NDL training pipeline. LightGCN produces user and item embeddings from the user--item graph. Two parallel branches estimate per-interaction noise: an \textbf{RCD branch} that scores the coherence between the current interaction representation $\mathbf{r}_{ui} = \mathbf{e}_u \odot \mathbf{e}_i$ and the user's historical centroid $\bar{\mathbf{r}}_u$ (active signal $\alpha^{\text{RCD}}_{ui}$, independent of training error), and a \textbf{Loss branch} that consumes the EMA-smoothed CE loss ($\alpha^{\text{loss}}_{ui}$, complementary in cold-start, exploration, and multi-interest cases). The two scores are blended by \textbf{activity-aware fusion}---weighting RCD higher for users with rich history and the loss signal higher for cold-start users---and the fused estimate $\alpha_{ui}$ drives a per-interaction dimension-adaptive mask: a suspicious interaction is allowed only a fraction of the embedding dimensions rather than being discarded outright. A \textbf{contrastive regularization branch} (bottom) shapes the embedding geometry to make coherence more discriminative; it does not itself produce a noise score nor enter the fusion.

\subsection{Representation Coherence Detection (RCD)}
\label{sec:rcd}

\paragraph{Interaction Representation}
We define the interaction representation as the element-wise product of user and item embeddings:
\begin{equation}
    \mathbf{r}_{ui} = \mathbf{e}_u \odot \mathbf{e}_i \in \mathbb{R}^d.
    \label{eq:interaction_repr}
\end{equation}
This representation captures each embedding dimension's contribution to user--item affinity.

\paragraph{Historical Mean Representation}
The user's historical mean representation aggregates all past interactions:
\begin{equation}
    \bar{\mathbf{r}}_u = \frac{1}{|\mathcal{N}_u|} \sum_{j \in \mathcal{N}_u} \mathbf{e}_u \odot \mathbf{e}_j \in \mathbb{R}^d.
    \label{eq:hist_mean}
\end{equation}
This vector serves as a compact summary of the user's typical interaction pattern in embedding space. In practice, $\bar{\mathbf{r}}_u$ is recomputed at the beginning of each training epoch using current embeddings to reflect the latest representation state.

\paragraph{Coherence Score}
The coherence between the current interaction and the historical pattern is measured via cosine similarity:
\begin{equation}
    \text{coh}_{ui} = \frac{\mathbf{r}_{ui} \cdot \bar{\mathbf{r}}_u}{\|\mathbf{r}_{ui}\| \cdot \|\bar{\mathbf{r}}_u\| + \varepsilon} \in [-1, 1],
    \label{eq:coherence}
\end{equation}
where $\varepsilon > 0$ (set to $10^{-8}$) is a small constant for numerical stability.

\paragraph{RCD Noise Estimate}
Lower coherence indicates greater deviation from the user's established pattern, suggesting higher noise probability:
\begin{equation}
    \alpha^{\text{RCD}}_{ui} = \sigma\left(\frac{\tau_{\text{coh}} - \text{coh}_{ui}}{\tau_{\text{coh}}}\right) \in (0, 1),
    \label{eq:rcd_alpha}
\end{equation}
where $\sigma(\cdot)$ is the sigmoid function and $\tau_{\text{coh}}$ is a temperature parameter. When $\text{coh}_{ui} = \tau_{\text{coh}}$, $\alpha^{\text{RCD}}_{ui} = \sigma(0) = 0.5$, i.e., the model remains neutral about that interaction. Smaller $\tau_{\text{coh}}$ makes the sigmoid steeper, increasing RCD's sensitivity to coherence differences. Because $\text{coh}_{ui}$ is computed from the existing embedding geometry rather than from prediction outcomes, RCD can flag suspicious interactions \emph{before} they trigger any loss feedback---this is what we mean by ``active.''

\subsection{Multi-Signal Fusion and Dimension Adaptation}
\label{sec:fusion}

CE loss provides a complementary \emph{passive} signal: hard-to-fit interactions are more likely noisy, giving $\alpha^{\text{loss}}_{ui} = \sigma(\tilde{\ell}_{ui}/\tau_{\text{loss}}) \in (0,1)$ on the EMA-smoothed CE loss $\tilde{\ell}_{ui}$ with temperature $\tau_{\text{loss}}$. RCD detects geometric deviation in embedding space, while CE loss detects prediction-level fitting difficulty---the two are complementary. For active users with rich histories $\bar{\mathbf{r}}_u$ is reliable and RCD is trustworthy; for cold-start users $\bar{\mathbf{r}}_u$ is unstable and the loss signal should weigh more. We thus fuse the two with an activity-aware weight:
\label{sec:activity_weighting}
\begin{align}
    \alpha_{ui} &= w_u \alpha^{\text{RCD}}_{ui} + (1-w_u)\alpha^{\text{loss}}_{ui}, \label{eq:fusion} \\
    w_u &= \min\!\left(\frac{|\mathcal{N}_u|}{\delta_{\text{act}}},\,1\right), \label{eq:activity_weight}
\end{align}
where $\delta_{\text{act}}$ is the activity threshold hyperparameter.

The fused noise estimate $\alpha_{ui}$ is then converted to an effective dimension count $d_{ui} = \lfloor (1 - \alpha_{ui}) d \rfloor$ and a binary mask $m_{ui}(k) = \mathds{1}[k \leq d_{ui}]$, $k=1,\ldots,d$. The prediction score is
\begin{equation}
    \hat{y}_{ui} = (\mathbf{e}_u \odot \mathbf{e}_i \odot \mathbf{m}_{ui}) \cdot \mathbf{1} = \sum_{k=1}^{d_{ui}} e_u(k)\,e_i(k).
    \label{eq:masked_pred}
\end{equation}
Noisy interactions ($\alpha_{ui}$ large $\Rightarrow d_{ui}$ small) participate with fewer dimensions, limiting their contamination of the embedding space, while clean ones use the full capacity. The mechanism is fine-grained and interpretable: $d_{ui}$ directly reveals the model's per-interaction noise assessment.

\subsection{Contrastive Regularization}
\label{sec:cl}

To further stabilize coherence estimation, NDL introduces a lightweight contrastive regularization module. Inspired by SimCSE~\cite{gao2021simcse} and InfoNCE-driven contrastive frameworks in vision~\cite{chen2020simclr,he2020moco}, we deliberately depart from graph-structural perturbation (e.g., SimGCL's uniform-noise scheme~\cite{yu2022simgcl}): for each mini-batch we take the LightGCN embeddings of the unique users $\mathcal{B}_u$ and unique items $\mathcal{B}_i$ and apply two independent dropouts to each side to obtain $L_2$-normalized augmented views. 

We optimize the symmetric InfoNCE~\cite{oord2018cpc} loss $\mathcal{L}_{\text{CL}}$ with cosine similarity, temperature $\tau_{\text{cl}}$, and in-batch unique users/items as negatives---keeping the regularizer at $O(|\mathcal{B}|^2)$ per batch and outside the noise-aware forward pass.

The CL module's sole purpose is to give RCD a more uniform, more discriminative embedding space to operate on. The final training objective is:
\begin{equation}
    \mathcal{L} = \mathcal{L}_{\text{CE}} + \lambda \cdot \mathcal{L}_{\text{CL}},
    \label{eq:total_loss}
\end{equation}
where $\lambda$ controls the CL strength. When $\lambda = 0$, the framework degenerates to the core NDL variant without CL; when $\lambda > 0$, we obtain the full NDL.

\section{Experiments}
\label{sec:experiments}

\subsection{Experimental Setup}

\paragraph{Datasets and Protocol.}
We use three public benchmarks: Yelp\footnote{\url{https://www.yelp.com/dataset}} (business), Amazon-book\footnote{\url{http://jmcauley.ucsd.edu/data/amazon/}} (e-commerce), and MovieLens-1M\footnote{\url{https://grouplens.org/datasets/movielens/1m/}} (movie). Following the noisy-training/clean-testing protocol of~\cite{wang2021denoising,lai2025dar}, interactions with rating $<3$ are treated as noise and only included in training; validation/test sets contain only clean ($\geq 3$) interactions. We report Recall@$K$ and NDCG@$K$ for $K\in\{5,10,20,50,100\}$, taking test metrics at the best-validation epoch to avoid leakage. 

The three benchmarks span two orders of magnitude in sparsity: Yelp contains 45{,}548 users, 57{,}396 items, and 1{,}672{,}520 interactions (sparsity 0.0640\%); Amazon-book contains 80{,}464 users, 98{,}663 items, and 2{,}714{,}021 interactions (sparsity 0.0342\%); and MovieLens-1M contains 6{,}040 users, 3{,}953 items, and 706{,}391 interactions (sparsity 2.9586\%).

\begin{table}[H]
\centering
\caption{NDL hyperparameter settings. CL-related parameters $\lambda=0.001$, $\tau_{\text{cl}}=0.2$, dropout$=0.1$ are fixed across datasets.}
\label{tab:hyperparams}
\begin{tabular}{lccc}
\toprule
Hyperparameter & MovieLens & Yelp & Amazon-book \\
\midrule
$\tau_{\text{coh}}$ & 0.1 & 0.5 & 0.5 \\
$\tau_{\text{loss}}$ & 5.0 & 10.0 & 5.0 \\
$\delta_{\text{act}}$ & 50 & 30 & 25 \\
$\gamma$ (EMA) & 0.9 & 0.9 & 0.5 \\
\bottomrule
\end{tabular}
\end{table}

\paragraph{Baselines.}
We focus on explicit denoising methods sharing our noisy-training/clean-testing protocol; self-supervised methods such as SGL~\cite{wu2021sgl} and SimGCL~\cite{yu2022simgcl} target overall recommendation under random splits without noise distinction and are not directly comparable. Baselines: \textbf{LightGCN}~\cite{he2020lightgcn} (backbone, no denoising), \textbf{T-CE} and \textbf{R-CE}~\cite{wang2021denoising}, \textbf{UDT}~\cite{chua2024unified}, and \textbf{DAR}~\cite{lai2025dar} (SIGIR 2025, SOTA).

\begin{table*}[!t]
\centering
\caption{Main results on three datasets. Best results are in \textbf{bold}; second-best are \underline{underlined}. The percentage shows relative improvement over DAR, with \textit{p}-values from two-sided \textit{t}-test for significance testing.}
\label{tab:main_results}
\begin{tabular}{l|l|cccccc}
\toprule
Dataset & Method & R@5 & R@10 & R@20 & N@5 & N@10 & N@20 \\
\midrule
\multirow{8}{*}{\rotatebox{90}{Yelp}}
& LightGCN       & 0.0229 & 0.0391 & 0.0655 & 0.0232 & 0.0286 & 0.0368 \\
& T-CE        & 0.0244 & 0.0422 & 0.0698 & 0.0247 & 0.0306 & 0.0392 \\
& R-CE        & 0.0227 & 0.0387 & 0.0657 & 0.0230 & 0.0283 & 0.0368 \\
& UDT        & 0.0270 & 0.0456 & 0.0769 & 0.0277 & 0.0337 & 0.0435 \\
& DAR        & \underline{0.0292} & \underline{0.0498} & \underline{0.0827} & \underline{0.0297} & \underline{0.0365} & \underline{0.0467} \\
\cmidrule(lr){2-8}
& NDL (Ours)        & \textbf{0.0328} & \textbf{0.0548} & \textbf{0.0889} & \textbf{0.0335} & \textbf{0.0405} & \textbf{0.0512} \\
& Improv.    & +12.33\% & +10.04\% & +7.50\% & +12.79\% & +10.96\% & +9.64\% \\
& \textit{p}-value    & 1.97e-05 & 1.52e-05 & 7.04e-06 & 5.39e-06 & 6.44e-06 & 5.70e-06 \\
\midrule
\multirow{8}{*}{\rotatebox{90}{Amazon-book}}
& LightGCN       & 0.0187 & 0.0313 & 0.0513 & 0.0177 & 0.0219 & 0.0280 \\
& T-CE        & 0.0217 & 0.0358 & 0.0582 & 0.0209 & 0.0255 & 0.0323 \\
& R-CE        & 0.0189 & 0.0315 & 0.0518 & 0.0179 & 0.0221 & 0.0282 \\
& UDT        & 0.0252 & 0.0414 & 0.0674 & 0.0240 & 0.0293 & 0.0371 \\
& DAR        & \underline{0.0273} & \underline{0.0449} & \underline{0.0721} & \underline{0.0260} & \underline{0.0317} & \underline{0.0399} \\
\cmidrule(lr){2-8}
& NDL (Ours)        & \textbf{0.0322} & \textbf{0.0507} & \textbf{0.0794} & \textbf{0.0300} & \textbf{0.0360} & \textbf{0.0447} \\
& Improv.    & +17.95\% & +12.92\% & +10.12\% & +15.38\% & +13.56\% & +12.03\% \\
& \textit{p}-value    & 1.62e-07 & 1.52e-07 & 9.79e-08 & 1.03e-07 & 1.58e-07 & 9.32e-08 \\
\midrule
\multirow{8}{*}{\rotatebox{90}{MovieLens-1M}}
& LightGCN       & 0.1017 & 0.1642 & 0.2543 & 0.3813 & 0.3464 & 0.3347 \\
& T-CE        & 0.0993 & 0.1612 & 0.2507 & 0.3773 & 0.3426 & 0.3309 \\
& R-CE        & 0.1017 & 0.1641 & 0.2539 & 0.3837 & 0.3473 & 0.3349 \\
& UDT        & 0.1047 & 0.1690 & 0.2622 & 0.3858 & 0.3511 & 0.3410 \\
& DAR        & \underline{0.1066} & \underline{0.1723} & \underline{0.2645} & \underline{0.3913} & \underline{0.3566} & \underline{0.3460} \\
\cmidrule(lr){2-8}
& NDL (Ours)        & \textbf{0.1126} & \textbf{0.1838} & \textbf{0.2802} & \textbf{0.4078} & \textbf{0.3730} & \textbf{0.3616} \\
& Improv.    & +5.63\% & +6.67\% & +5.94\% & +4.22\% & +4.60\% & +4.51\% \\
& \textit{p}-value    & 2.05e-07 & 7.69e-09 & 1.04e-08 & 6.62e-08 & 6.56e-09 & 4.80e-09 \\
\bottomrule
\end{tabular}
\end{table*}

\paragraph{Implementation Details.}
All methods use LightGCN with $d=32$, 2-layer propagation, learning rate $10^{-3}$. NDL hyperparameters per dataset are in Table~\ref{tab:hyperparams}: $\delta_{\text{act}}$ is set near each dataset's median user-interaction count so that the RCD/loss crossover (Eq.~\ref{eq:activity_weight}) tracks the actual active/cold boundary; $\tau_{\text{coh}}$ is more conservative on sparser datasets ($0.5$ vs.\ $0.1$), where noisier centroids favor a softer sigmoid in Eq.~(\ref{eq:rcd_alpha}).

All experiments are conducted on a single NVIDIA A10 GPU.

\subsection{Main Results}

Table~\ref{tab:main_results} presents the main results. NDL outperforms all baselines across all datasets on Recall@$K$ ($K = 5, 10, 20$) and NDCG@$K$ ($K = 5, 10, 20$), with consistent improvements across both metrics, indicating that NDL's gains stem from systematic improvement in overall ranking quality.

The strongest baseline DAR already opens a comfortable margin over LightGCN, UDT, and the others, so the additional gap that NDL opens over DAR cannot be attributed to extra capacity---all methods share the same LightGCN backbone and embedding dimensionality. The gain must come from the orthogonal active coherence signal.

A second pattern is that NDL's relative advantage grows with data sparsity. In sparse settings the noisy fraction of each user's history is larger, and methods that cannot tell signal from noise let small amounts of bad interactions drag the user embedding off course; RCD, detecting the deviation directly in embedding space, contains this drift even when clean signal is scarce. Since long-tail users with sparse history dominate real-world recommendation traffic, this is also where the gain matters most in practice.

\begin{table}[H]
\centering
\caption{Ablation study on MovieLens-1M. Each row removes one component from the full NDL; $\Delta$ reports the relative change in Recall@5 against the full model.}
\label{tab:ablation}
\setlength{\tabcolsep}{3pt}
\begin{tabular}{lcccc}
\toprule
Variant & Recall@5 & Recall@10 & Recall@20 & $\Delta$ (vs. Full) \\
\midrule
Full NDL & 0.1126 & 0.1838 & 0.2802 & -- \\
$-$ RCD & 0.1003 & 0.1675 & 0.2651 & $-$10.9\% \\
$-$ CL & 0.1071 & 0.1698 & 0.2559 & $-$4.9\% \\
$-$ Loss signal & 0.1084 & 0.1744 & 0.2651 & $-$3.7\% \\
$-$ Activity fusion & 0.1103 & 0.1785 & 0.2722 & $-$2.0\% \\
\bottomrule
\end{tabular}
\end{table}

\subsection{Ablation Study}

We conduct comprehensive ablation studies to verify the necessity of each component. Results are shown in Table~\ref{tab:ablation}. Key findings:

\begin{itemize}[leftmargin=*]
    \item \textbf{RCD ($-$10.9\%)}: The most critical component. Without RCD, the model can only rely on the loss signal for reliability estimation, with performance approaching the LightGCN baseline (0.1003 vs.\ 0.1017), demonstrating the crucial value of the active representation coherence signal.
    \item \textbf{CL ($-$4.9\%)}: Removing CL yields the core NDL variant. This result shows that on relatively dense datasets like MovieLens where DAR is already strong, RCD alone can improve top-$K$ recommendations, but CL is needed to improve representation geometry for stable gains across more ranking metrics.
    \item \textbf{Loss signal ($-$3.7\%)}: Removing the loss signal hurts performance, confirming the complementarity of the two signal sources. However, RCD contributes more ($-$10.9\% vs.\ $-$3.7\%): RCD carries the primary denoising signal, while Loss serves as a supplement for low-activity users and the representation-unstable phase.
    \item \textbf{Activity fusion ($-$2.0\%)}: Adaptive weighting provides consistent improvement through differentiated treatment of active and cold-start users.
\end{itemize}

\subsection{Noise Detection Analysis}

A unique advantage of NDL is that its noise estimates are interpretable and directly evaluable. In this section, we use the noise labels available in the evaluation protocol (interactions with ratings $< 3$ are noisy) to quantitatively assess the quality of the RCD signal.

\subsubsection{Clean--Noisy Ranking Separation}

Aggregate Recall hides whether NDL's improvement comes from genuinely promoting relevant items or from uniformly lifting all positives. To separate the two, we split test positives by rating into clean (rating $\geq 3$) and noisy (rating $< 3$), and look at how they are separated inside the Top-$K$ list.

\begin{figure}[H]
    \centering
    \begin{subfigure}[t]{\columnwidth}
        \centering
        \includegraphics[width=\linewidth]{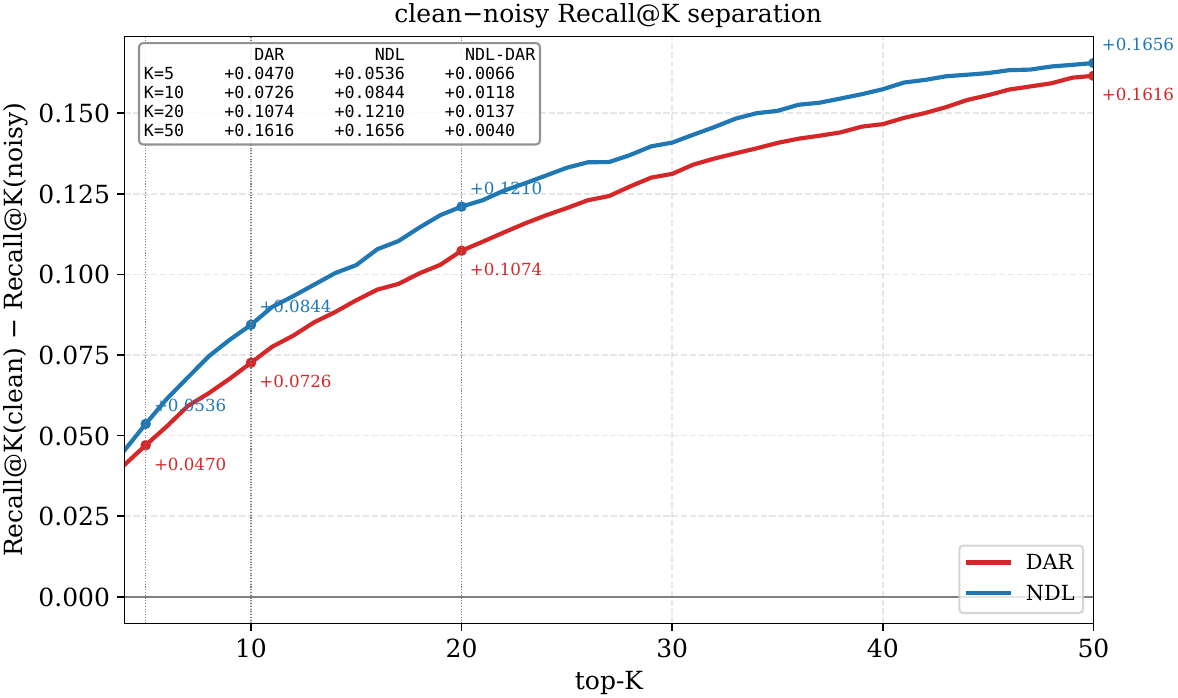}
        \caption{Clean--noisy Recall@$K$ separation at the best validation checkpoint across $K\in[5,50]$. NDL stays above DAR over the full range.}
        \Description{Line plot comparing the clean--noisy Recall@K separation score, defined as Recall@K on clean test positives minus Recall@K on noisy test positives, for NDL and DAR over K from 5 to 50 on the MovieLens validation set. NDL (blue line) lies consistently above DAR (red line) across the full K range, with the largest gap of roughly 0.0118 to 0.0137 occurring at K equal to 10 and 20, indicating that NDL ranks clean (high-rated) positives above noisy (low-rated) positives more strongly than DAR.}
        \label{fig:separation}
    \end{subfigure}

    \vspace{0.6em}

    \begin{subfigure}[t]{\columnwidth}
        \centering
        \includegraphics[width=\linewidth]{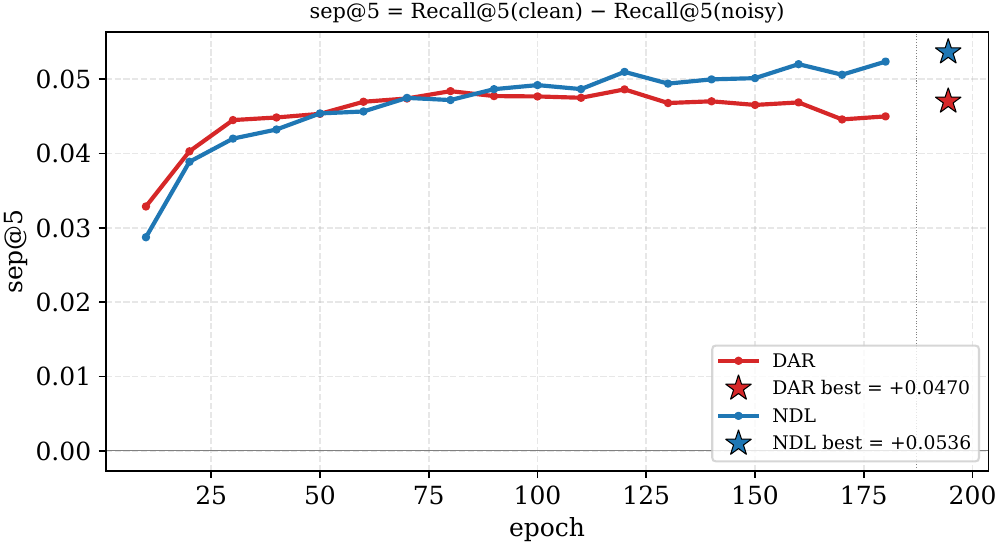}
        \caption{Training evolution of $\text{sep}@5$. NDL keeps widening the gap after DAR plateaus, indicating a stable training effect rather than an isolated checkpoint.}
        \Description{Two line plots tracking the clean--noisy Recall@K separation of NDL and DAR across training epochs on the MovieLens validation set. DAR's separation peaks early (around epoch 80) and then plateaus or slightly regresses. NDL's separation continues to grow, crossing above DAR around epoch 50 to 60 and remaining strictly above DAR for every subsequent epoch until training ends, indicating a stable widening of the clean--noisy gap rather than an isolated lucky checkpoint.}
        \label{fig:separation_evolution}
    \end{subfigure}
    \caption{Clean--noisy separation analysis on MovieLens validation. $\text{sep}@K = \text{Recall}@K(\text{clean}) - \text{Recall}@K(\text{noisy})$, where clean and noisy are test positives with rating~$\geq 3$ and $<3$, respectively. Higher values mean the model more strongly ranks clean positives above noisy ones in the Top-$K$ list.}
    \label{fig:sep_combined}
\end{figure}

Figure~\ref{fig:separation} compares clean–noisy separation between NDL and DAR across K. Across the entire $K \in [5, 50]$ range, NDL (blue) consistently exceeds DAR (red), leading by +0.0066 / +0.0118 / +0.0137 / +0.0040 at $K = 5, 10, 20, 50$ respectively. Since Recall@$K$ over all positives is strictly a weighted average of its clean and noisy components, the comprehensive dominance of sep@$K$ directly demonstrates that NDL's Recall@$K$ improvement over DAR does not come from ``uniformly lifting all positive samples'' but from \emph{preferentially promoting relevant (high-rated) items}. This advantage is most pronounced in the $K = 10$--$20$ range that recommendations practically care about, with NDL achieving approximately 13--16\% relative improvement in separation over DAR.

Figure~\ref{fig:separation_evolution} shows the evolution of this advantage throughout training. DAR's clean--noisy separation peaks in early training (around epoch 80) and then plateaus or slightly regresses, indicating model saturation on the training objective. In contrast, NDL continues to widen the separation after DAR stops improving, surpassing DAR around epoch 50--60 and maintaining growth until training completion. This demonstrates that NDL's denoising mechanism continuously guides the optimizer toward solutions that ``more firmly rank clean items first,'' rather than merely converging to a slightly sharper version of DAR. Notably, after the crossover epoch, NDL is never overtaken---every subsequent checkpoint's sep@$5$ is strictly greater than DAR's---indicating a stable training effect rather than an accidental fluctuation at a particular epoch.

\subsubsection{Clean--Noisy Embedding Separability}

\begin{figure}[H]
    \centering
    \includegraphics[width=\columnwidth]{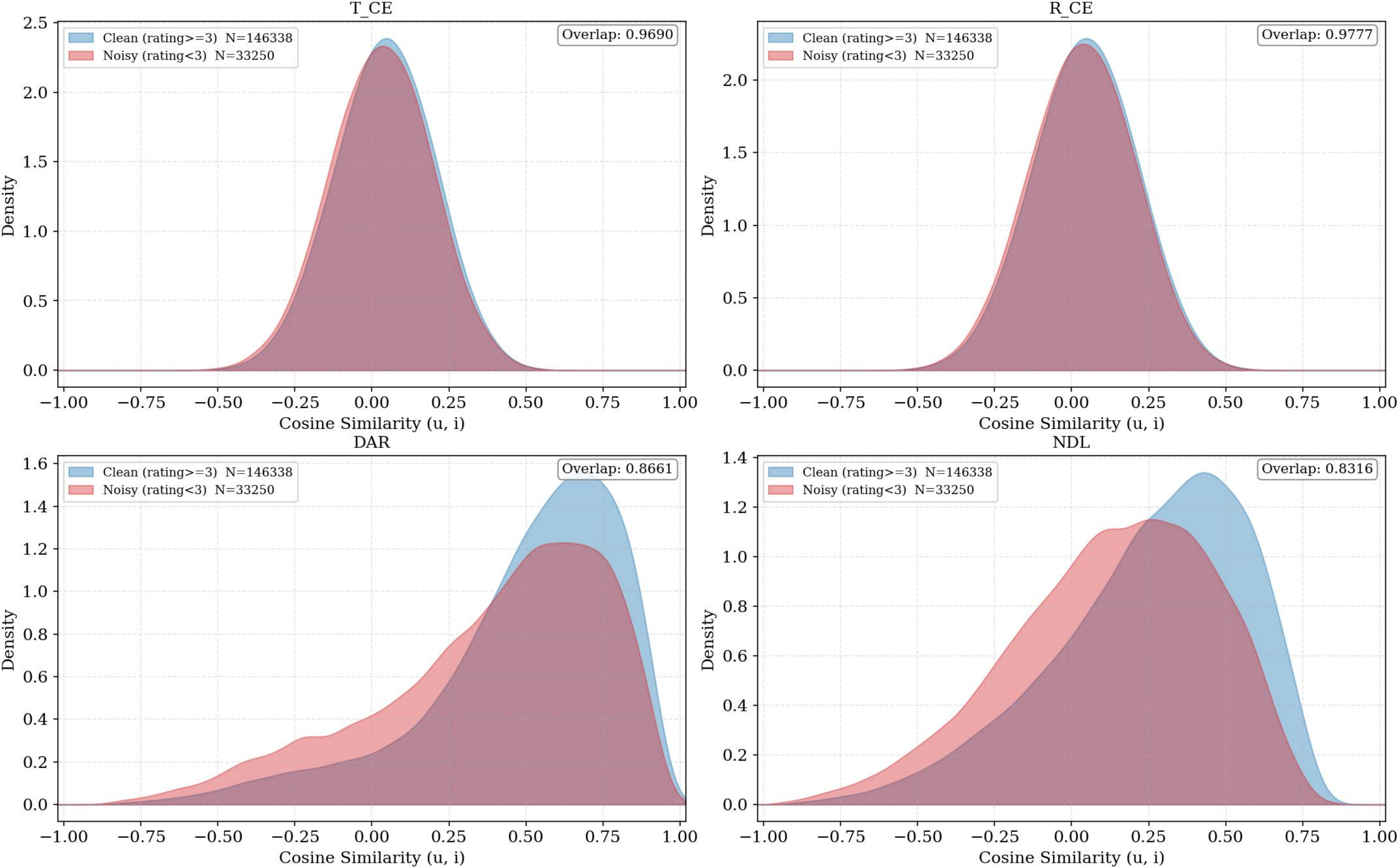}
    \caption{Clean vs.\ noisy embedding cosine distribution comparison. Overlap is the minimum envelope integral of the two KDE curves---lower indicates better separation.}
    \Description{Four kernel density plots comparing the user--item embedding cosine similarity distributions of clean interactions (rating at least 3, 146{,}338 samples, blue) and noisy interactions (rating below 3, 33{,}250 samples, red) under four denoising methods on the MovieLens validation set. R-CE has the highest distribution overlap of 0.9777 with both curves nearly coinciding around cosine 0.05. T-CE has overlap 0.9690 with similarly compressed curves. DAR has overlap 0.8661 with both peaks clustered in the high similarity region around 0.5 to 0.8. NDL has the lowest overlap of 0.8316 with the noisy peak at cosine about 0.05 and the clean peak at cosine about 0.25, showing a clear bimodal shift. The progression confirms that NDL produces the most discriminative embedding geometry between clean and noisy interactions.}
    \label{fig:cosine_dist}
\end{figure}

Figure~\ref{fig:cosine_dist} shows the ``clean vs.\ noisy'' embedding cosine distribution comparison across different denoising methods on the MovieLens validation set. The $x$-axis is the user--item embedding cosine similarity $\cos(\mathbf{e}_u, \mathbf{e}_i)$ for each interaction; the $y$-axis is kernel density estimation. Blue represents clean interactions (rating $\geq 3$, $N = 146{,}338$), and red represents noisy interactions (rating $< 3$, $N = 33{,}250$). Overlap is defined: $\text{Overlap} = \int \min(p_{\text{clean}}(x), p_{\text{noise}}(x))\, dx \in [0, 1]$; lower values indicate easier separation between the two classes in embedding space.

The ordering R-CE (0.9777) $>$ T-CE (0.9690) $>$ DAR (0.8661) $>$ NDL (0.8316) is fully consistent with the methodological progression: from ``only modifying training gradients'' (T-CE/R-CE) $\to$ ``loss-based dimension adaptation'' (DAR) $\to$ ``representation-coherence-based dimension adaptation'' (NDL). NDL achieving the lowest overlap demonstrates that RCD's geometric noise signal directly translates into a more discriminative embedding space, and this embedding-level separability improvement is the fundamental reason for NDL's consistent superiority over DAR and DenoisingRec baselines on downstream Recall@$K$ metrics.

\subsection{Cold-Start Analysis}

The ``active/cold-start'' split in this subsection is for visualization only and is independent of the fusion threshold $\delta_{\text{act}}$ in \S\ref{sec:activity_weighting}.

\begin{figure}[H]
    \centering
    \includegraphics[width=\columnwidth]{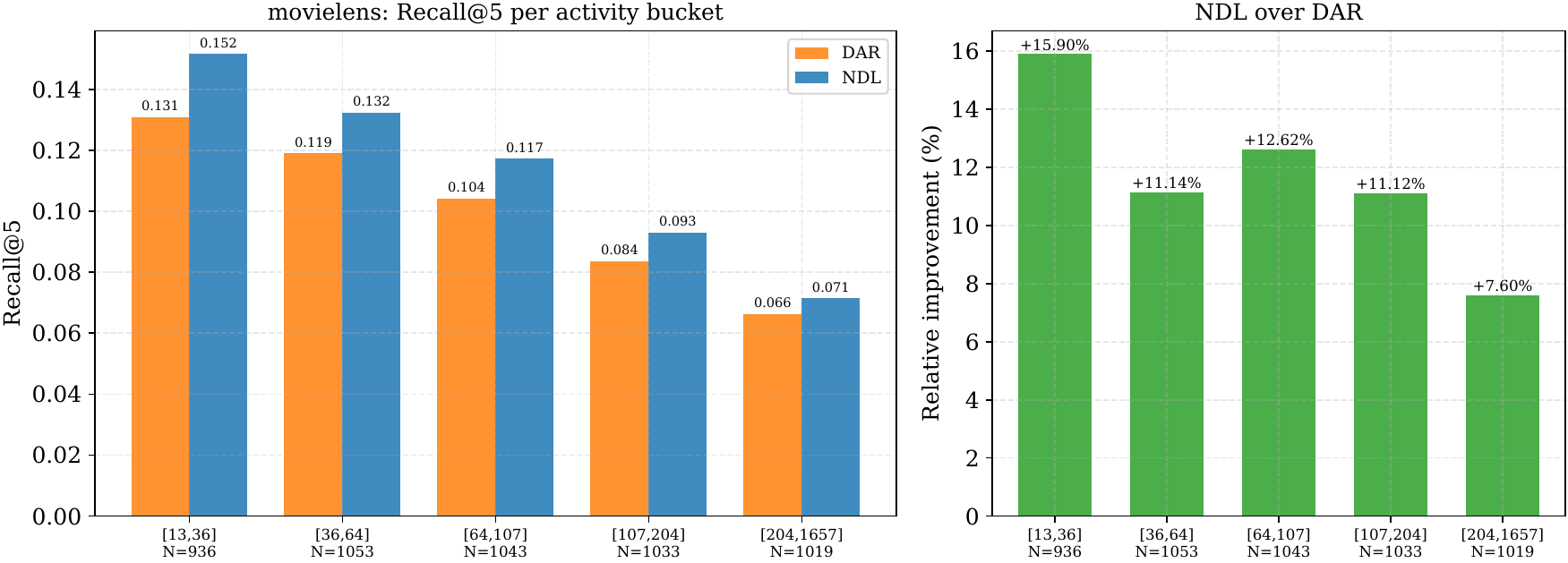}
    \caption{Binned Recall@5 improvement by user activity level on MovieLens-1M test set. Users are equally divided into 5 bins by training interaction count. \textbf{Left}: Absolute Recall@5 per bin. \textbf{Right}: Relative improvement (NDL $-$ DAR) / DAR, generally decreasing from +15.90\% for the sparsest bin to +7.60\% for the most active bin.}
    \Description{Two-panel bar chart on the MovieLens-1M test set, where users are equally divided into five bins by training interaction count, with about 1{,}000 users per bin. The left panel shows the absolute Recall@5 of NDL (blue) and DAR (orange) for each activity bin, with NDL above DAR in every bin. The right panel shows the relative improvement of NDL over DAR per bin, generally decreasing from +15.90 percent in the sparsest bin to +7.60 percent in the most active bin, indicating the largest gains come from cold users with sparse interactions.}
    \label{fig:bucket_recall}
\end{figure}

As shown in Figure~\ref{fig:bucket_recall}, NDL consistently outperforms DAR across all user activity levels, but the magnitude of improvement is not uniform: gains are largest for the sparsest users and gradually shrink as user activity increases. This pattern indicates that the proposed denoising mechanism is most beneficial for cold users---where each interaction carries disproportionate weight and a single noisy edge can mislead preference modeling---while gains naturally saturate for active users who already have sufficient clean signal to average out noise.

\subsection{Hyperparameter Sensitivity Analysis}

We analyze NDL's sensitivity to three key hyperparameters on MovieLens, each corresponding to a core design: coherence temperature $\tau_{\text{coh}}$ (noise estimation), activity threshold $\delta_{\text{act}}$ (activity-aware fusion), and CL weight $\lambda$ (representation geometry). For $\tau_{\text{coh}} \in \{0.1, 0.3, 0.5, 0.7, 1.0\}$, performance remains stable when $\tau_{\text{coh}} \leq 0.5$ but degrades sharply beyond 0.5 (NDCG drops $\sim$4.4\% at $\tau_{\text{coh}}=1.0$) as $\alpha^{\text{RCD}}_{ui}$ approaches a constant; we recommend $\tau_{\text{coh}}=0.1$ with a robust range $[0.1, 0.3]$. For $\delta_{\text{act}} \in \{20, 50, 70, 100, 150\}$, $\delta_{\text{act}}=50$ achieves peak performance at $K \geq 10$, while smaller values let cold users with unreliable $\bar{\mathbf{r}}_u$ inject errors and larger values force moderately active users onto loss signals, diluting RCD's benefits; this optimum aligns with MovieLens's median activity level. For $\lambda \in \{0, 0.0001, 0.001, 0.01\}$, Recall@5 reaches its peak at $\lambda=0.001$ (0.1126), while $\lambda=0.01$ triggers premature convergence (0.1045) and $\lambda=0.0001$ is suboptimal (0.1098); $\lambda=0.001$ consistently performs best across all three datasets and is set as the default.

\section{System Launch Design and Deployment Path}
\label{sec:deployment}

NDL targets noise patterns common in Netease News feed ($\sim$20M users, $\sim$8M items), mainly accidental clicks and quick-exit interactions. 

\paragraph{Integration.} Our online retrieval system employs multiple candidate generators, among which a LightGCN-based graph collaborative filtering model serves as one of the primary channels—the same backbone NDL extends. At serving time the model outputs only the trained user/item embeddings, identical in shape and storage to the current LightGCN deployment, so NDL adds \textbf{no online latency, parameters, or memory} and requires no change to ranking or serving.

\paragraph{Training cost.} Per-user centroids $\bar{\mathbf{r}}_u$ are refreshed once per epoch from messages LightGCN already aggregates. Per-interaction coherence is the same as the dot-product prediction. Extra training-time state is bounded: a centroid table and a per-interaction EMA-loss buffer, both of which stay offline and are not shipped to serving. In our offline benchmarks, end-to-end training wall-time grew by \textbf{approximately 4.2\%} over the LightGCN baseline---negligible at production scale.

\paragraph{Deployment Status and Path.} NDL has passed offline evaluation on production-scale data and internal code review. The deployment path follows our team's standard launch protocol: (1)~\emph{shadow mode} traffic replay for consistency verification, (2) ~\emph{5\% A/B} traffic, and (3)~\emph{full ramp} with online metrics (\textbf{CTR and dwell, stratified on long-tail/cold users}) monitoring. The model has not yet been deployed online; the launch is scheduled for an upcoming release cycle of the feed recommender.


\section{Conclusion}
\label{sec:conclusion}

We identify a blind spot in implicit-feedback denoising: prior methods all read noise from the same passive loss signal, inheriting its training-stage and memorization ambiguities. Representation coherence—a purely geometric quantity computable before any prediction error—provides an orthogonal active signal that rescues noise the loss misses and protects clean interactions the loss wrongly suppresses. Fusing the two via activity-aware weighting consistently beats either alone, with gaps widening on sparser data. Moreover, mapping the fused estimate to a per-interaction effective dimension count yields finer control than gradient reweighting or edge dropping, and plugs into any graph-CF backbone. While effective, our approach assumes short-term preference stability and incurs minor per-epoch centroid computation overhead. Handling users with strong multi-interest patterns remains an interesting future direction.

\section*{GenAI Usage Disclosure}
\label{sec:genai_disclosure}

The authors used ChatGPT and Gemini for grammar polishing and auxiliary script generation; all AI outputs were verified, revised, and approved by the human authors.

\bibliographystyle{ACM-Reference-Format}
\bibliography{refs}

\end{document}